\documentstyle[epsf]{article}

\newcommand{\bee}{\begin{equation}}
\newcommand{\ee}{\end{equation}}
\newcommand{\beea}{\begin{eqnarray}}
\newcommand{\eea}{\end{eqnarray}}
\newcommand{\rme}{{\rm e}}

\begin{document}
\thispagestyle{empty}
\parskip=12pt
\raggedbottom

\def\mytoday#1{{ } \ifcase\month \or
 January\or February\or March\or April\or May\or June\or
 July\or August\or September\or October\or November\or December\fi
%\space\number\day ,
 \space \number\year}
\noindent
\hspace*{9cm} BUTP--95/14\\
\hspace*{9cm} COLO-HEP-361\\
\vspace*{1cm}
\begin{center}
{\LARGE The classically perfect fixed point action for SU(3) gauge 
theory}\footnote{Work supported in part by Schweizerischer Nationalfonds,
NSF Grant PHY-9023257 
and U.~S. Department of Energy grant DE--FG02--92ER--40672}

\vspace{1cm}

Thomas DeGrand,
Anna Hasenfratz \\
Department of Physics \\
 University of Colorado,
Boulder CO 80309-390 

\vspace{.5cm}

Peter Hasenfratz, 
Ferenc Niedermayer\footnote{On leave from the Institute of Theoretical
Physics, E\"otv\"os University, Budapest}
\\
Institute for Theoretical Physics \\
University of Bern \\
Sidlerstrasse 5, CH-3012 Bern, Switzerland

\vspace{0.5cm}

\mytoday \\ \vspace*{0.5cm}

\nopagebreak[4]

\begin{abstract}
In this paper (the first of a series) we describe the construction
of fixed point actions for lattice $SU(3)$ pure gauge theory.
Fixed point actions have scale invariant instanton solutions
and the spectrum of their quadratic part is exact (they are classical perfect
actions).
We argue that the fixed point action is even 1--loop quantum perfect,
i.e. in its physical predictions there are no $g^2 a^n$ cut--off 
effects for any $n$. 
We discuss the construction of fixed point operators and present
examples. The lowest order $q {\bar q}$ potential $V(\vec{r})$ obtained
from the fixed point Polyakov loop correlator is free
of any cut--off effects which go to zero  as an inverse
power of the distance $r$.
\end{abstract}

\end{center}
\eject

\section{Introduction and summary}

Using a discrete space--time lattice as an ultraviolet regulator
for a quantum field theory
 also opens the way to many non--perturbative
methods. At the same time however the lattice
introduces various artifacts like rotational symmetry
breaking. 
 Cut--off independent continuum quantities are obtained
in the limit of zero lattice spacing. Depending on the relevance
of the lattice artifacts this limit can be reached on relatively
coarse lattices or one might need lattices where the lattice
spacing in physical units is extremely small.
The main problem for
numerical lattice calculations is the  control of the  finite
lattice spacing effects.

The relevance of the lattice artifacts on the numerical results
can strongly
depend on the specific regularization  or lattice action.
Most numerical calculations  to date of QCD use the Wilson plaquette
action. This action is the simplest
 one which can be used but it has some undesirable properties.
For example, asymptotic scaling sets in slowly and non--smoothly.
Even more important is that
the continuum limit is reached at lattice spacing less
then 0.1 fm
and, consequently,  nowadays typical pure gauge or quenched
calculations are done on spatial volumes $32^3$
or larger \cite{WEINGARTEN}.

Two of us \cite{HN} suggested a radical solution, to use a perfect
lattice action which is completely free of lattice artifacts.
That such a perfect action exists follows from Wilson's
renormalization group theory \cite{sigma5}. 
The fixed point (FP) of a renormalization group (RG) transformation
and the renormalized trajectory (RT)
along which the action runs under repeated RG transformations
form a perfect quantum action. The FP action itself reproduces
all the important properties of the continuum classical action
(it is a classical perfect action). Although at finite coupling values
the FP action is not ``quantum perfect'' \footnote{In some limiting cases
the FP action might be trivially related to the renormalized trajectory
 \cite{BIETENHOLZ}.}
it is expected to be
a very good first approximation.
This program was successfully
carried out for the 2--dimensional non--linear $\sigma$ model. An
action parametrized by about 20 parameters was found that showed
no finite lattice spacing effects even at correlation length $\xi
\sim 3$ lattice spacings. The computational overhead to simulate this action
(about a factor of three relative to the nearest neighbor action)
is negligible compared to the gain obtained by the great reduction
of lattice artifacts.

\begin{figure}[htb]
\begin{center}
%\vskip 10mm
\leavevmode
\epsfxsize=90mm
\epsfbox{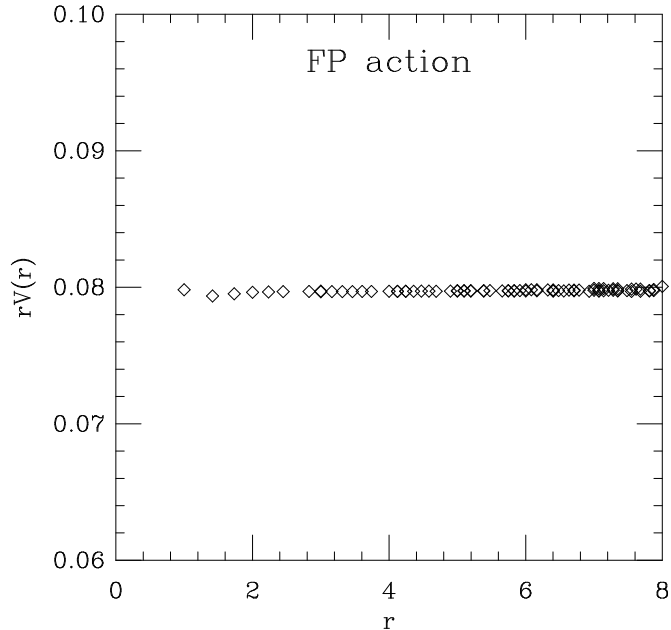}
%\vskip 10mm
\end{center}
\caption{ The quantity $rV(r)$ constructed from the correlators of the
FP Polyakov loops and using the FP action.
Note that $1/4\pi=.0796$. }
\label{fig:rvrperfect}
\end{figure}

\begin{figure}[htb]
\begin{center}
%\vskip 10mm
\leavevmode
\epsfxsize=90mm
\epsfbox{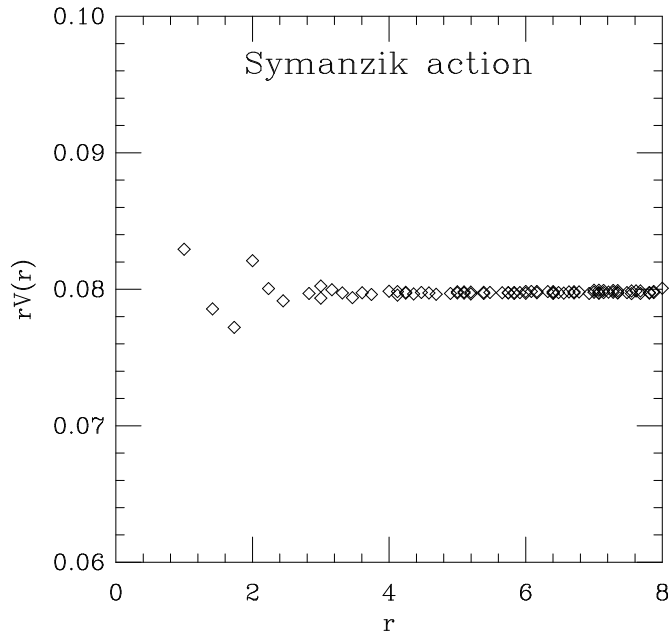}
%\vskip 10mm
\end{center}
\caption{ The quantity $rV(r)$ constructed
for the Symanzik action.
}
\label{fig:rvrsyman}
\end{figure}

\begin{figure}[htb]
\begin{center}
%\vskip 10mm
\leavevmode
\epsfxsize=90mm
\epsfbox{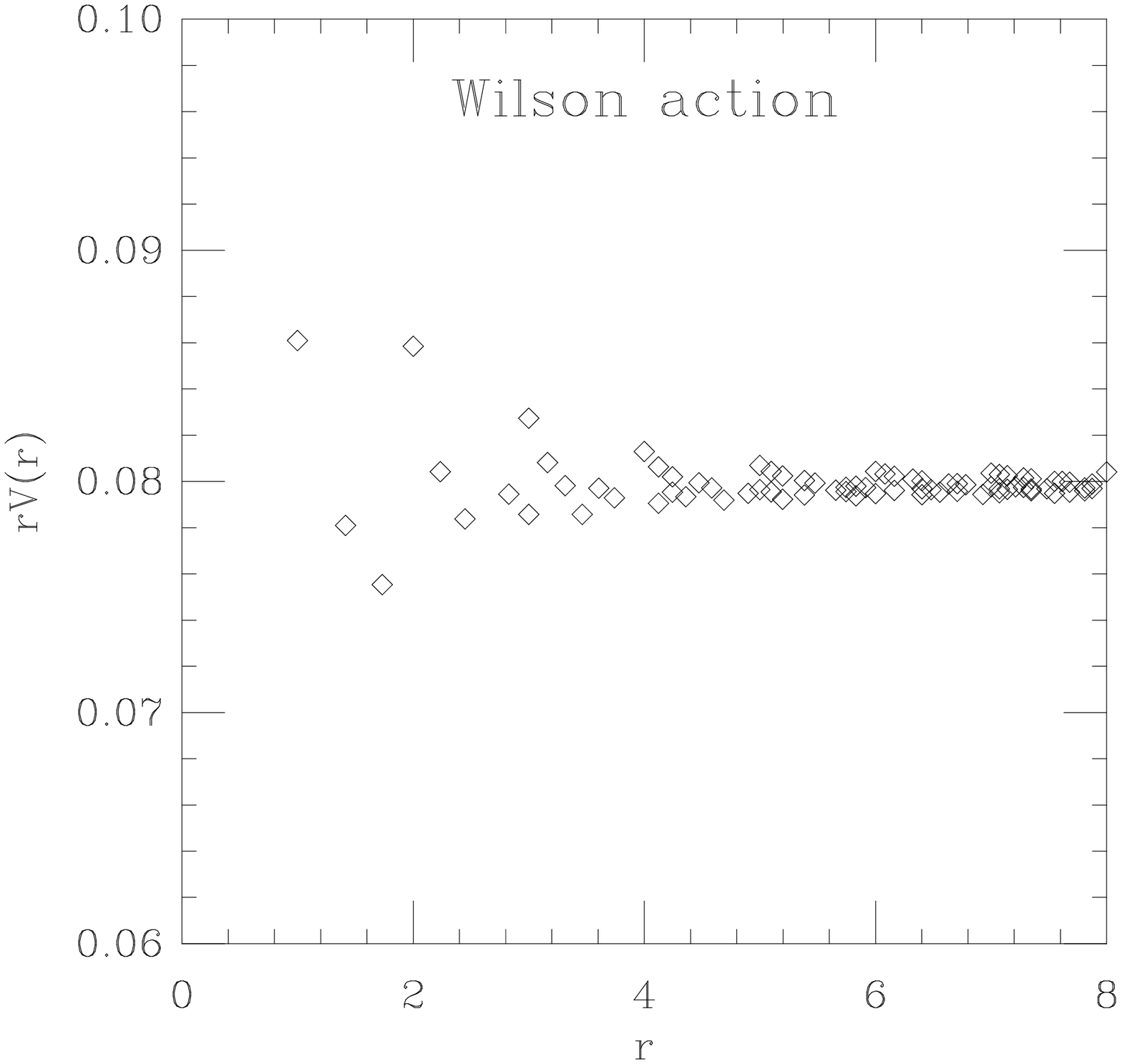}
%\vskip 10mm
\end{center}
\caption{ The quantity $rV(r)$ constructed
for the Wilson action.
}
\label{fig:rvrwilson}
\end{figure}

In this and a follow up paper \cite{II} we report on a similar project
for the 4--dimensional pure gauge SU(3) theory. We construct
FP actions and FP operators and investigate their properties.
In this first paper the basic equations are presented
and the properties of the FP action and operators are investigated
on fields with small fluctuations. The methods used are mostly analytic.
In the second paper we turn to the problem of using the FP action
in numerical simulations on coarse lattices. A simple parametrization
is derived and the scaling properties of the action are studied
by measuring torelon masses and the potential.

In the present paper we consider two different type of RG
transformations. Both transformations contain free parameters
which are used to make the FP action as short ranged as possible.
For any practical application this is a basic requirement. 
The FP action is defined by a classical saddle point equation.
It follows from this equation that the FP action has exact stable
instantons. The value of the action for the instantons
is $8\pi^2$ independently of the size of the instanton
(scale invariance). The FP equation can be solved analytically
on fields with small fluctuations by expanding in powers of 
the vector potentials.
The quadratic part of the FP action describes massless gluons
with an exact relativistic spectrum: $E=| \vec{p} |$,
$p_i \in (-\infty,\infty)$.
This is to be contrasted with that of the Wilson action where
this dispersion relation is valid only for $|\vec{q}| \ll 1$ and,
in addition, the spectrum is restricted to
the Brillouin zone $|q_i| \le \pi$.
For small $| \vec{q}|$ the Symanzik tree level improved action
\cite{SYMANZIK}, \cite{WEISZ}
performs better, but then, in the middle of the Brillouin zone
it breaks down completely producing complex energies.

An exciting question is how the FP action performs in 1--loop
perturbation theory. We present a formal RG argument that the
physical predictions are free of cut--off effects O($a^n$)
and O($g^2 a^n$) for any $n$ provided the size of the system 
is larger then the range of the action.
Is therefore the classical perfect
action automatically 1--loop quantum perfect?
Our RG arguments are formal, not rigorous.
The statement is supported further by a detailed 1--loop
calculation in the non--linear $\sigma$ model \cite{AHNP}
where  the cut--off effects of the mass gap are studied in a finite
box of size $L$. The power like decaying cut--off effects 
$(a/L)^n$ were found
several orders of magnitude smaller (and consistent with zero
within the errors of
the calculation) than those in the standard or
in the tree level improved Symanzik model.

We need not only perfect actions but perfect Polyakov, or Wilson
loops, currents, etc. as well. As a first step we discuss here
how to construct FP operators. We derive the explicit form of the
FP Polyakov loop on fields with small fluctuations and obtain 
the corresponding
tree level $q \bar q$ potential $V(\vec{r})$.
We show that $V(\vec{r})$ is free of any cut--off effects
which decay as a power $r^{-k}$ for large $r$. 
This property is to be contrasted with that of the standard and Symanzik
action. In the latter case  only the leading O($a^2$) distortion 
is canceled. The situation is illustrated in 
Figs.~\ref{fig:rvrperfect},
\ref{fig:rvrsyman} and \ref{fig:rvrwilson}.

Kerres, Mack, and Palma\cite{MACK} have recently described the
construction of a perfect action for scalar electrodynamics in four 
dimensions and at finite temperature.
Some of the basic ideas of \cite{MACK} overlap with those introduced
in \cite{HN} and in the present work.

The outline of the paper is as follows.
In Section 2 the two different RG transformations we considered 
are defined. In Section 3 the FP equation is derived and the
statement about classical solutions of the FP action is discussed
briefly. In Sections 4 and 5 the FP action is constructed at the
quadratic level and the spectrum is studied. In Section 6 the main steps
leading to the cubic terms of the action are presented.
Section 7 deals with
the question of whether the FP action is 1--loop quantum perfect.
The problem of FP operators is discussed in Section 8.

\section{The RG transformation}

Consider an SU(N) gauge theory defined on a hypercubic lattice.
The partition function is
\bee
Z = \int DU e^{-\beta S(U)}  ,                                 \label{1}
\ee
where $\beta S(U)$ is some lattice representation of the continuum
action. It is a function of the  products of link variables
$U_\mu(n)=e^{iA_\mu(n)} \in$SU(N) along
arbitrary closed loops.
The normalization is fixed such that on smooth configurations
the action takes the standard continuum form
\bee
\beta S(U) \rightarrow {1 \over {2g^2}} \int d^4x 
Tr \left( F_{\mu\nu} F_{\mu\nu} \right) ,                       \label{2}
\ee
where
\bee
\beta={2N \over g^2}.                                            \label{3}
\ee

\begin{figure}[htb]
\begin{center}
\vskip 10mm
\leavevmode
\epsfxsize=80mm
\epsfbox{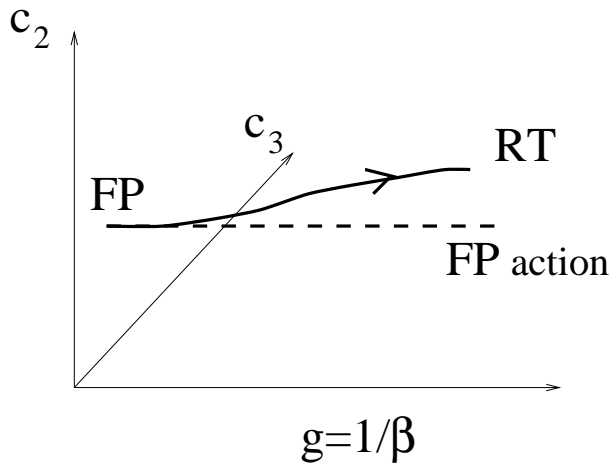}
\vskip 10mm
\end{center}
\caption{Schematic flow diagram for asymptotically free gauge
theories.}
\label{fig:rgflow}
% FIGURE_FILE rgflow.ax
\end{figure}

Let us denote the couplings of $\beta S(U)$ by 
$\beta$, $c_2,c_3,\ldots$. 
The action can be represented by a point in this infinite dimensional 
coupling constant space as shown in fig. \ref{fig:rgflow}.

Under repeated real space renormalization group transformations
the action moves in this multidimensional parameter space.
We shall consider RG transformations with a scale factor of 2.
The blocked link variable $V_{\mu}(n_B) \in$SU(N), which lives on the 
coarse lattice with lattice unit $a'=2a$, 
is coupled to a local average of the original link variables.
The new action is defined as
\bee
\rme^{-\beta' S'(V)} = \int DU 
\exp\left\{ -\beta \left( S(U)+T(U,V)\right) \right\}  ,        \label{4}
\ee
where the blocking kernel $T(U,V)$ is taken in the form
\bee
T(U,V)= \sum_{n_B,\mu} \left\{ -{\kappa \over N}
{\rm Re Tr} \left( V_{\mu}(n_B) Q^{\dagger}_{\mu}(n_B)\right) 
+ {\cal N}\left( Q_{\mu}(n_B) \right) \right\} .                \label{5}
\ee

The $N \times N$ complex matrix $Q_{\mu}(n_B)$ is an average
of the fine link variables in the neighbourhood of the coarse link 
$l_B=(n_B,\mu)$. Its explicit form will be specified later.
The last term in Eq.~(\ref{5}) assures that the partition
function remains invariant under the RG transformation,
\bee
\rme^{\beta {\cal N}(Q)} = \int dW 
\exp\left\{ \beta {\kappa \over N}
{\rm Re Tr}(W Q^{\dagger}) \right\}, ~~~ W \in {\rm SU(N)} \,. \label{6}
\ee
The parameter $\kappa$ is free and will be used for optimization
as in \cite{HN}.
We require the averaging procedure to be gauge invariant
\bee
T \left( U^g, V^g \right) = T(U,V),                           \label{7}
\ee
where $U^g$ and $V^g$ are gauge transformed configurations
on the fine and coarse lattice, respectively.
This property, together with the gauge symmetry of the action $S(U)$,
assures that $S'(V)$ is also gauge invariant.

In this paper we consider two different block transformations
which will be referred to as type~I and type~II, respectively.
The first transformation we used is a modified version
of the Swendsen transformation \cite{RNG} 
shown in fig.~\ref{fig:type1} where $c$, the relative weight
of the staples vs. the central link, is a tunable parameter:

\begin{figure}[htb]
\begin{center}
\vskip 10mm
\leavevmode
\epsfxsize=80mm
\epsfbox{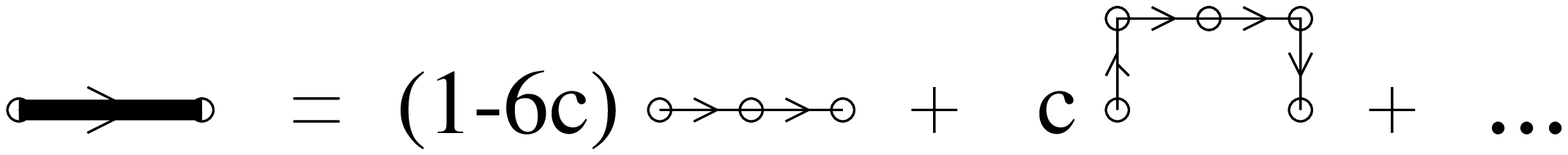}
%\vskip 10mm
\end{center}
\caption{The renormalization group transformation of type~I.}
\label{fig:type1}
\end{figure}

$$
Q_\mu(n_B) =  (1-6c)U_\mu(n)U_\mu(n+\mu) + ~~~~~~~~~ 
$$
\bee
 c \sum_{\nu \ne \mu} \left[
U_\nu(n)U_\mu(n+\nu)U_\mu(n+\mu+\nu)U^\dagger_\nu(n+2\mu)  + \right.
                                                               \label{8}
\ee
$$
 \left.
U_\nu^\dagger(n-\nu)U_\mu(n-\nu)U_\mu(n+\mu-\nu)U_\nu(n+2\mu-\nu)
 \right] . \nonumber
$$

%\beea
%& Q_\mu(n_B)  =  (1-6c)U_\mu(n)U_\mu(n+\mu) +  ~~~~~~~~~~~~~~~~~~~~~~~~~
%  \nonumber \\
% & ~~~~~~~~ c \sum_{\nu \ne \mu} \left[
%  U_\nu(n)U_\mu(n+\nu)U_\mu(n+\mu+\nu)U^\dagger_\nu(n+2\mu) + \right. ~~~~
%                                                       \label{8} \\
% & ~~~~~~~~~~~~~~~~~~~~~~~~  \left.
%U_\nu^\dagger(n-\nu)U_\mu(n-\nu)U_\mu(n+\mu-\nu)U_\nu(n+2\mu-\nu)
% \right] . \nonumber
%\eea

\begin{figure}[htb]
\begin{center}
\vskip 10mm
\leavevmode
\epsfxsize=80mm
\epsfbox{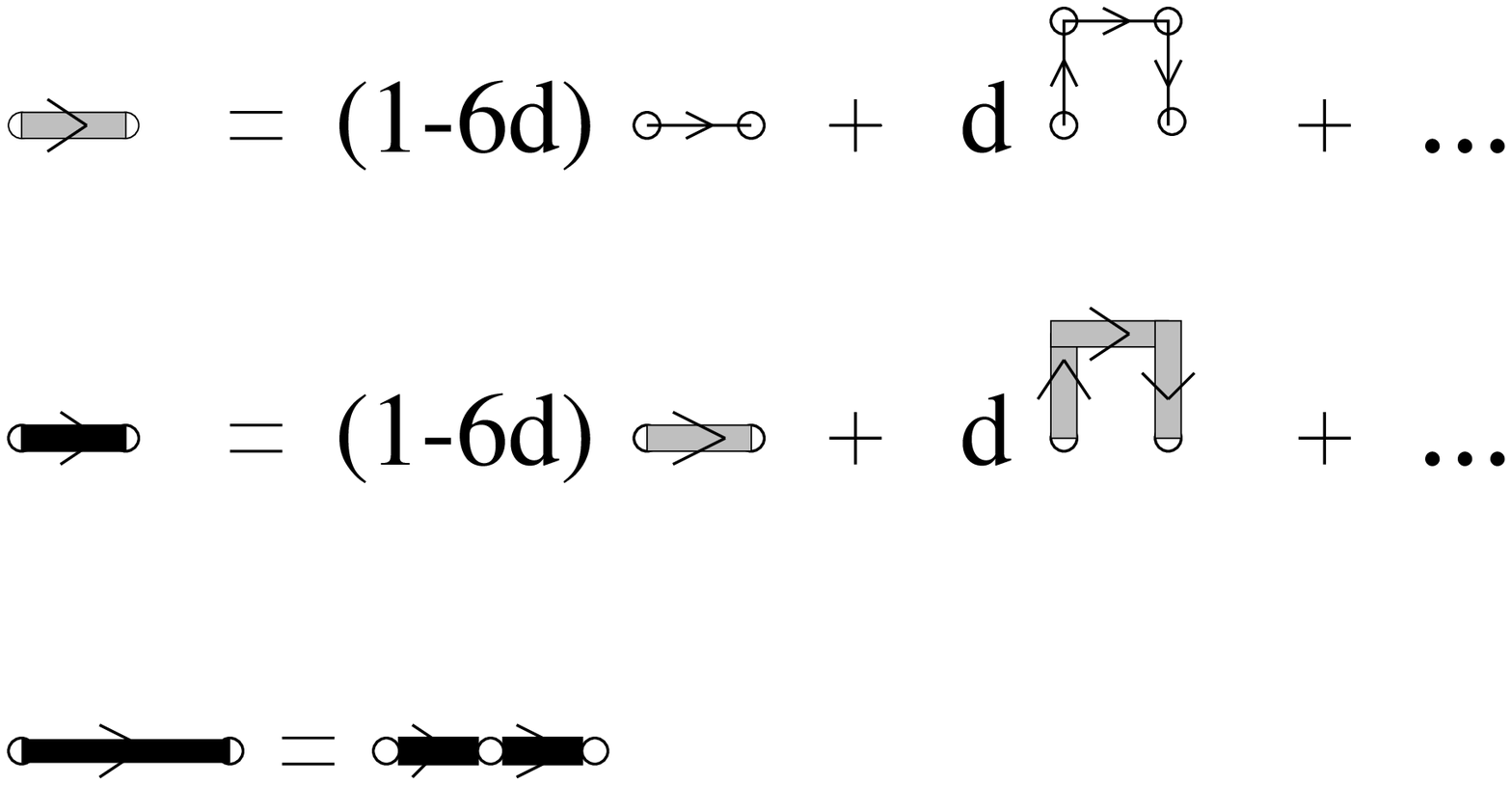}
%\vskip 10mm
\end{center}
\caption{The renormalization group transformation of type~II.}
\label{fig:type2}
\end{figure}

The second transformation uses smearing
to create fuzzy links. The matrix $Q_{\mu}(n_B)$
is the product of the two fuzzy link variables obtained
after two smearing steps.
The construction is shown in fig.~\ref{fig:type2}.

Here $c$ and $d$ are tunable parameters. With $c=1/7$ the 
transformation Eq.~(\ref{8}) is identical to the one used
in the $\beta$ function calculations earlier \cite{RNG}.

It is expected that these RG transformations have a fixed point (FP)
on the $\beta = \infty$ critical surface. The FP has one marginally
relevant and infinitely many irrelevant directions. The trajectory
which leaves the FP along the marginal direction is the renormalized
trajectory (RT).

As it was argued in \cite{HN} the FP and the RT describe perfect actions.
There are no lattice artifacts along the RT. For reasons discussed in
the Introduction and later, we shall refer to the FP action $\beta S^{FP}
(U)$ as the classical perfect action. This is the action we shall
consider in this work. Although this action does not lie on the RT at
finite $\beta$ values, it has amazingly nice properties at large $\beta$ 
 and shows no, or tiny, cut--off effects even on very coarse
configurations.

\section{The fixed point equation}

On the critical surface $\beta \to \infty$
Eq.~(\ref{4}) reduces to a saddle point problem
giving
\bee
S^{FP}(V)=\min_{ \{U\} } \left( S^{FP}(U) +T(U,V)\right),  \label{9}
\ee
where $S^{FP}$ is the FP action.
For a given coarse configuration $\{ V \}$ there are 
infinitely many minimizing $\{ U \}$ configurations
which differ by gauge transformations leaving the
averages $Q_{\mu}(n_B)$ invariant. The value of the
FP action $S^{FP}(V)$ is not influenced by this
freedom, however.

Using Eqs.~(\ref{5},\ref{6}) the saddle point equation
can be written as
\bee
S^{FP}(V)=\min_{\{ U \} } \left\{
S^{FP}(U)-{\kappa\over N} 
\sum_{n_B,\mu} \left[ {\rm Re Tr} \left( V_{\mu}(n_B)
Q_{\mu}^{\dagger}(n_B)\right) -f\left( Q_{\mu}(n_B) \right)
\right] \right\},                                             \label{10}
\ee
where
\bee
 f(Q)=\max_{W} \left\{ {\rm Re Tr} (W Q^{\dagger}) \right\},
 ~~~~ W \in {\rm SU}(N) .  \label{10a}
\ee

This equation has a very similar structure to the one in the 
$d=2$ non--linear $\sigma$--model \cite{HN} and its implications
for the classical solutions of $S^{FP}$ are the same:
if the configuration $\{ V \}$ satisfies the FP classical 
equations and  is a local minimum of $S^{FP}(V)$
then the configuration $\{ U(V) \}$ on the finer lattice
minimizing the r.h.s.. of Eq.~(\ref{10}) satisfies the FP
equations as well.
In addition, the value of the action remains unchanged,
$S^{FP}(V)=S^{FP}(U(V))$.

Since the proof goes along the same lines as in the $d=2$
$\sigma$--model \cite{HN}, we can be brief here.
If $\{ V \}$ is a solution of the classical equations of motion
$\delta S^{FP}/\delta V =0$, then Eq.~(\ref{10}) implies that
on the minimizing configuration $\{ U(V)\}$ the expression
in square brackets in Eq.~(\ref{10}) takes its
maximum value, i.e. zero.
Then it follows that
\bee
\left. {\delta S^{FP}(U) \over \delta U } 
\right|_{ U=U(V) } =0,                                      \label{11}
\ee
and $S^{FP}(V)=S^{FP}(U(V))$.

According to this result the FP action has exact scale invariant 
instanton solutions with action $=8\pi^2$ as in the continuum 
theory. This property allows  one to define a correct topological 
charge on the lattice \cite{INSTANTON}. In this paper, however,
we will not follow this problem further.

\section{The FP action at the quadratic level}

The FP equation Eq.~(\ref{10}) is valid for any configuration 
$\{V\}$, be it smooth or rough. It is a highly non--trivial 
equation. On $\{V\}$ configurations with small fluctuations, however
Eq.~(\ref{10}) can be expanded in powers of the vector
potentials and studied analytically.

On the quadratic level we can write\footnote{We suppress
the index ``FP''.}.
\bee
2N S(U)=\sum_{n,r}\rho_{\mu \nu}(r){\rm Tr} \left(
A_{\mu}(n+r)A_{\nu}(n)\right) +{\rm O}\left( A^3\right)         \label{12}
\ee
$$
{\phantom{2N S(U)}} = {1\over V} \sum_k 
\tilde{\rho}_{\mu \nu}(k) {\rm Tr}
\left( \tilde{A}_{\mu}(-k)\tilde{A}_{\nu}(k) \right)
+{\rm O}\left( \tilde{A}^3\right),
$$
where $\rho_{\mu \nu}(r)$ (in momentum space
$\tilde{\rho}_{\mu \nu}(k)$) are the quadratic couplings
to be determined\footnote{In the following we suppress the tilde
for Fourier transformed quantities.}.
Gauge symmetry and the normalization
condition Eq.~(\ref{2}) give the constraints
\beea
&\widehat{k}_{\mu}^{*} \rho_{\mu \nu}(k)=0,               \label{13} \\
& \rho_{\mu \nu}(k) \to k^2 \delta_{\mu \nu} -
k_{\mu} k_{\nu} ~~~{\rm for~} k\to 0,
\nonumber
\eea
where we introduced the notation 
$\widehat{k}_{\mu}=\rme^{i k_{\mu}}-1$.
For both RG transformations (fig.~\ref{fig:type1} and fig.~\ref{fig:type2})
$Q_{\mu}(n_B)$ is a sum over products of $U$ matrices
along paths connecting the points $n$ and $n+2\widehat{\mu}$.
It is easy to show that
\bee
{\rm Re Tr} \left( V_{\mu}(n_B) Q_{\mu}^{\dagger}(n_B)\right)
-f(Q_{\mu}(n_B)) = -{1\over 2}{\rm Tr} \left(
\Gamma_{\mu}(n_B)-B_{\mu}(n_B)\right)^2
 + {\rm O}\left( {\rm cubic} \right),                         \label{14}
\ee
where
\bee
\Gamma_{\mu}(n_B)= \sum_{r,\nu} 
\omega_{\mu \nu}(2n_B -r) A_{\nu}(r) \,.                      \label{15}
\ee
In Eq.~(\ref{15}) the tensor $\omega_{\mu \nu}$ is
fixed by the form of the RG transformation kernel in
fig.~\ref{fig:type1} and fig.~\ref{fig:type2}:
\bee
\omega_{\mu \nu}(k)=\left( 1+\rme^{ik_{\mu}}\right)
\left[ \delta_{\mu \nu} \left( 1- c(\widehat{k}\widehat{k}^*) \right)
+ c \widehat{k}_{\mu} \widehat{k}_{\nu}^* \right],
~~~{\rm type~I},                                              \label{16a}
\ee
\bee
\omega_{\mu \nu}(k)=\left( 1+\rme^{ik_{\mu}}\right)
\left[ \delta_{\mu \nu} \left( 1- d(\widehat{k}\widehat{k}^*) \right)^2
+ d \left( 2-d(\widehat{k}\widehat{k}^*)\right)
 \widehat{k}_{\mu} \widehat{k}_{\nu}^* \right],
~~{\rm type~II}.                                              \label{16b}
\ee
At the quadratic level the recursion relation has the form
$$
{1 \over V_B} \sum_{k_B} \rho'_{\mu \nu}(k_B)
{\rm Tr}\left( B_{\mu}(-k_B)B_{\nu}(k_B)\right) =
$$
\bee
\min_{ \{ A \} } \left\{ {1\over V} \sum_k
\rho_{\mu \nu}(k)
{\rm Tr}\left( A_{\mu}(-k)A_{\nu}(k)\right) \right. +
                                                             \label{17}
\ee
$$
\left.
 \kappa {1 \over V_B} \sum_{k_B} {\rm Tr}
\left[ \left( \Gamma_{\mu}(-k_B)-B_{\mu}(-k_B)\right)
\left( \Gamma_{\mu}(k_B)-B_{\mu}(k_B)\right) \right]
\right\} \,,
$$
where $V_B={1 \over 16} V$ is the volume of the blocked lattice. 
To proceed further we have
to introduce a temporary gauge fixing to get rid of the remaining
gauge freedom at the sites of the fine lattice not representing
a site on the coarse lattice.
One obtains then from Eq.~(\ref{17}) a recursion relation
for the inverse of $\rho$:
\bee
D'_{\mu \nu}(q_B)={1\over 16} \sum_{l=0}^1
\left[ \omega({q_B\over 2} +\pi l)
D({q_B \over 2} +\pi l)
\omega^{\dagger}({q_B \over 2} +\pi l)
\right]_{\mu \nu} +{1\over \kappa} \delta_{\mu \nu} \,,      \label{18}
\ee
where $l=(l_0,l_1,l_2,l_3)$ is an integer vector, the summation
goes over $l_{\mu}=0,1$ and
\bee
D_{\mu \nu}(q_B)=
\left( \rho^{-1}(q_B)\right)_{\mu \nu} \,.            \label{19}
\ee

A way to find the FP solution of Eq.~(\ref{18}) is to start from
a $D^{(0)}_{\mu \nu}(q)$ with some gauge fixing parameter
$\alpha$
\bee
D^{(0)}_{\mu \nu}(q)={1\over (\widehat{q}\widehat{q}^*) } \delta_{\mu \nu}
+\alpha {\widehat{q}_{\mu}\widehat{q}^*_{\nu} \over 
(\widehat{q}\widehat{q}^*)^2} \,,
                                                                   \label{20}
\ee
and iterate Eq.~(\ref{18}) to the FP solution
$D^{FP}_{\mu \nu}$. The inverse of $D^{FP}_{\mu \nu}$
has a smooth $\alpha \to \infty$ limit which corresponds to
removing the gauge fixing. The resulting 
$\rho^{FP}_{\mu \nu}(k)$ is transverse and satisfies
Eq.~(\ref{13}).

The program sketched above is not completely trivial.
The main steps and some intermediate results are collected in Appendix A.

The fixed point $\rho^{FP}_{\mu \nu}$ depends on the parameters of the
RG transformation $(\kappa,c)$, or $(\kappa,d)$ for the kernels of type~I
and type~II, respectively. Although none of the theoretical properties
of the FP action depend on these parameters, the range of 
$\rho^{FP}_{\mu \nu}(r)$ (the range of the interaction) changes 
significantly as these parameters are varied. It is vital in applications
that the interaction is short ranged. Using this requirement we
found that the following parameter values are close to optimal:
\beea
\kappa=12.0,~~~ & c=0.12,~~~ {\rm type~I} \,,          \label{21} \\
\kappa=10.5,~~~ & d=0.077,~~~ {\rm type~II} \,. \nonumber 
\eea
Using these values the significantly non--zero couplings lie in
the unit hypercube and $\rho^{FP}_{\mu \nu}(r)$ goes rapidly
to zero with the distance. The decrease seems to be as fast as
$\approx\exp(-3r)$. In Table \ref{tab:type1} and 
\ref{tab:type2} we collected some of the 
elements of $\rho^{FP}_{\mu \nu}(r)$ for the RG transformation
of type~I and II, respectively. Using these tables, plus the
cyclic permutation of the coordinates and the reflection
properties of $\rho_{\mu \nu}(r)$ (Appendix B), further elements can
be obtained.

\begin{table}
\begin{center}
\begin{tabular}{c c | c  c}      \hline
  $r$     &    $\rho_{00}(r)$  & $r$ &  $\rho_{10}(r)$ \\   \hline
0~0~0~1 & -0.6718  & 0~0~0~0 &    -0.8007       \\
0~0~0~2 &  0.02420 & 0~0~0~1 &    -0.03100       \\
0~0~0~3 & -0.00422 & 0~0~0~2 &    -0.00170       \\
1~0~0~0 & -0.05951 & 0~1~0~0 &     0.01129       \\
0~0~1~1 & -0.04281 & 0~1~0~1 &    -0.00208       \\
1~0~0~1 &  0.01614 & 1~0~1~1 &     0.00880       \\
0~1~1~1 & -0.02724 & 2~1~0~0 &    -0.00387       \\
\hline
\end{tabular}
\end{center}
\caption{Some of the elements of $\rho_{00}(r)$ and $\rho_{10}(r)$
for the RG transformation of type~I.}
\label{tab:type1}
\end{table}

\begin{table}
\begin{center}
\begin{tabular}{c c | c c}      \hline
  $r$     &    $\rho_{00}(r)$  & $r$ &  $\rho_{10}(r)$ \\   \hline
0~0~0~1 & -0.5872  & 0~0~0~0 &    -0.7777       \\
0~0~0~2 &  0.00769 & 0~0~0~1 &    -0.04813       \\
0~0~0~3 & -0.00035 & 0~0~0~2 &    -0.00081       \\
1~0~0~0 &  0.00467 & 0~1~0~0 &    -0.00102       \\
0~0~1~1 & -0.07964 & 0~1~0~1 &    -0.00063       \\
1~0~0~1 & -0.00298 & 1~0~1~1 &     0.00804       \\
0~1~1~1 & -0.02490 & 2~1~0~0 &    -0.00571       \\
\hline
\end{tabular}
\end{center}
\caption{Some of the elements of $\rho_{00}(r)$ and $\rho_{10}(r)$ 
for the RG transformation of type~II.}
\label{tab:type2}
\end{table}

\section{The spectrum of the quadratic FP action}

Although the quadratic FP action Eq.~(\ref{12}) lives on the lattice,
it describes transverse massless gluons with an exact relativistic
spectrum $E(\vec{k})=|\vec{k}|$, $k_i \in (-\infty,\infty)$.
This follows from the fact that the FP action represents in fact
an infinitely fine lattice, i.e. the continuum.
A similar property has been demonstrated earlier for the $d=2$
 non--linear $\sigma$--model \cite{HN} where this statement follows
trivially from an explicit closed representation of the FP propagator.
This property is to be contrasted with that of the standard or
Symanzik improved actions where the spectrum is distorted
by O($a^2$), or O($a^4$) cut--off effects, respectively.
A distorted spectrum implies distorted thermodynamics.
Large cut--off effects are seen in the free energy at high
temperatures both in perturbation theory and in numerical simulations 
\cite{CUTOFF}.

The spectrum is determined by the singularities of the propagator
$D_{\mu \nu}(q)$. Consider this propagator after $n$ RG steps
as given by Eq.~(A7) in Appendix A.
The $\sim \delta_{\mu \nu}$
part of $D^{(0)}_{\mu \nu}(q)$ gives
for large $n$ the term
\bee
\left[ \Omega^{(n)}\left({q+2\pi l \over 2^n}\right)
\Omega^{(n)\dagger} \left({q+2\pi l \over 2^n}\right)
\right]_{\mu \nu}
{1\over (q+2\pi l)^2} \,.                             \label{22}
\ee
The tensor $\Omega^{(n)}$ is a smearing function which builds up
the blocked link after $n$ RG steps out of the original link
variables. This is a local function in configuration space whose
Fourier transform is smooth, $\Omega^{(n)}_{\mu \nu}(q=0)=\delta_{\mu \nu}$.
Eq.~(\ref{22}) has a pole at $k_0= \pm i |\vec{k}|$, where
$k=q+2 \pi l$. For any given $\vec{q}$, $q_i\in (0,2\pi)$ the summation
over $l$ gives a tower of poles in $D_{\mu \nu}(q)$ which reproduce
the full relativistic spectrum of the continuum theory.

The terms proportional to $1/\kappa$ in Eq.~(A7) are constructed out of
$\Omega_{\mu \nu}$ and do not change this conclusion\footnote{This part of the
FP propagator is more complicated than the one in the non--linear
$\sigma$--model because with our block transformations a link
variable contributes to several block averages.}.
Similarly, the part of $D_{\mu \nu}$ proportional to the gauge fixing
parameter $\alpha$ is removed at the end and does not influence
the form of $\rho^{FP}(q)$ (cf. Appendix A).

\begin{figure}[htb]
\begin{center}
%\vskip 10mm
\leavevmode
\epsfxsize=100mm
\epsfbox{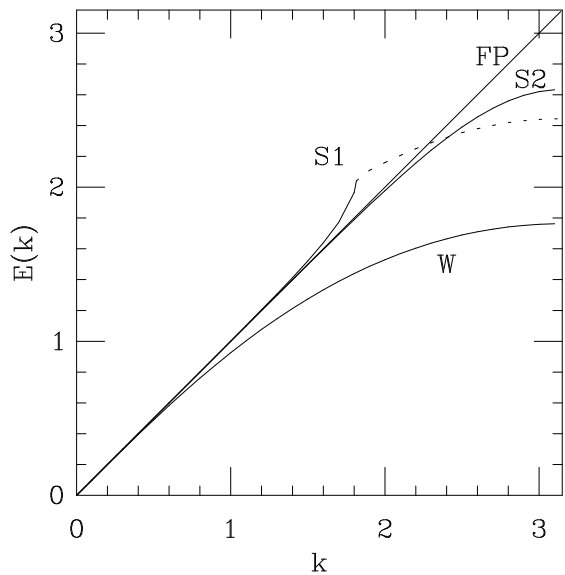}
%\vskip 10mm
\end{center}
\caption{The spectrum of the quadratic part for the Wilson action (W),
the Symanzik improved action for two different choices of the coefficients:
S1: $c_1=-1/12$, $c_2+c_3=0$, S2: $c_1=0$, $c_2+c_3=1/12$,
  and the FP action.}
\label{fig:spectrum}
\end{figure}

In Fig.~\ref{fig:spectrum} we compare the spectrum of the quadratic parts of
the Wilson, Symanzik improved and FP actions.
Our parameterization of the Symanzik actions follows the notation of
Weisz \cite{WEISZ}. For the Wilson and Symanzik actions the momentum in 
Fig.~\ref{fig:spectrum} is restricted to lie along one of the lattice
axes. The energy-momentum relation for the action labelled S1 becomes
complex at large $k$; the dotted line shows its real part.

\section{The connecting tensor $Z_{\mu \nu}(n)$ and the FP action 
at the cubic level}

In an SU(N) non Abelian theory gauge symmetry implies the presence
of terms in the action which are cubic in the vector potentials.
Gauge symmetry alone, however, does not fix the cubic part of the action.
There are combinations of Wilson loops whose contribution to the 
quadratic part is zero while to the cubic part is not.
The cubic part of the FP action is, however, uniquely given by
the FP equation Eq.~(\ref{10}).

One expands Eq.~(\ref{10}) up to terms cubic in the coarse
($B_{\mu}$) and fine ($A_{\mu}$) vector potentials.
The action is written in the form 
\bee
2N\cdot S(U)={1\over 2}\sum_{n,r} \rho_{\mu \nu}(r) 
A^a_{\mu}(n+r) A^a_{\nu}(n) +
                                                                \label{23}
\ee
$$
{1\over 12} f^{abc} \sum_{n,r_1,r_2} c_{\mu \nu \rho}(r_1,r_2)
A^a_{\mu}(n+r_1) A^b_{\nu}(n+r_2) A^c_{\rho}(n) + 
{\rm O}\left( A^4 \right),
$$
where in terms of the SU(N) generators,
\bee
\left[ T^a, T^b \right] = i f^{abc} T^c,
~~~{\rm Tr}T^a T^b={1\over 2} \delta_{ab} \,.                  \label{24}
\ee
The terms proportional to $\kappa$ in Eq.~(\ref{10}) are expanded to
cubic order as well. The quadratic and cubic coupling constant
tensors $\rho_{\mu \nu}(r)$ and $c_{\mu\nu\rho}(r_1,r_2)$ are determined 
then by the FP condition Eq.~(\ref{10}). 

After introducing a gauge fixing term in the action
and solving the quadratic problem one obtains the minimizing 
fine vector potentials as a function of the coarse fields (Appendix A).
In configuration space this relation can be written in the form
\bee
A^a_{\mu}(n)=\sum_{n'_B} Z_{\mu \rho}(n-2n'_B) B^a_{\rho}(n'_B).
                                                             \label{25}
\ee
Since the minimizing configuration is fixed only up to gauge
transformations (cf. Section 2), the very definition of the relation
in Eq.~(\ref{25}) requires gauge fixing. We used the type of gauge fixing
given in Eq.~(\ref{20}) and defined the connecting tensor
$Z_{\mu \nu}$ in the $\alpha\to\infty$ limit.
Using eqs.~(A5),(A6) one obtains
\bee
Z_{\mu\rho}(n-2n'_B)=\lim_{ {\alpha\to\infty \atop n\to\infty}}
{1\over V_B} \sum_{k_B} {1\over 16}\sum_{l=0}^1
\rme^{i ({1\over 2} k_B +\pi l)(n-2n'_B)} \times
                                                              \label{26}
\ee
$$
\left[ D^{(n)}({k_B\over 2}+\pi l)
\omega^{\dagger}({k_B\over 2}+\pi l)
\left( D^{(n+1)}(k_B)\right)^{-1} \right]_{\mu\rho}
$$
Note that sites on the fine lattice can be labeled as
 $n=2n_B+m$, $m_{\mu}=0,1$. In Fourier space one obtains

$$Z_{\mu\rho}(q)=\left[ D^{(n)}(q) \omega^{\dagger}(q) 
\left( D^{(n+1)}(2q) \right)^{-1} \right]_{\mu \rho} \,.
$$

The tensor $Z_{\mu \nu}$ is an important quantity which enters not only 
the cubic calculation, but also the construction of the fixed point
operators, like the FP Polyakov loop (cf. section~8).
The connecting function decays rapidly with  distance; the 
minimizing configuration is determined locally by the coarse vector
potential $B_{\mu}$. In Table \ref{tab:Z} a few leading
elements of $Z_{00}$ and $Z_{10}$ are collected for the two different
block transformations. A permutation of the coordinates and the
(somewhat unusual) reflection properties of $Z_{\mu \nu}$ (Appendix B)
help one to find its value for further related $\mu,\nu$ and $n$ values.

\begin{table}
\begin{center}
\begin{tabular}{c c c | c c c}      \hline
 $n$ & $Z_{00}^{I}$ & $Z_{00}^{II}$ &$n$& $Z_{10}^{I}$ & $Z_{10}^{II}$  \\  
 \hline
 0~0~0~0 & 0.4031  &  0.4442 & 0~0~0~0 &  0.0878 &  0.0751     \\
 0~1~0~0 & 0.1756  &  0.1413 & 0~1~0~0 & -0.0408 & -0.0475     \\
 0~0~1~1 & 0.0610  &  0.0623 & 2~1~0~1 &  0.0226 &  0.0153     \\ 
-1~1~0~0 & 0.0514  &  0.0526 & 3~1~0~0 &  0.0217 &  0.0148     \\
 0~1~1~1 & 0.0339  &  0.0345 &         &         &       \\
\hline
\end{tabular}
\end{center}
\caption{Some of the elements of the tensor $Z_{\mu \nu}$ which
gives the minimizing configuration in terms of the coarse vector
potential for the block transformation type~I and type~II.}
\label{tab:Z}
\end{table}

The minimizing vector potential is given by Eq.~(\ref{25})  only in the 
quadratic approximation. Nevertheless, this expression can also be used 
when we need the minimum in Eq.~(\ref{10}) up to cubic order. The error made
is only quartic. This way all the cubic terms in Eq.~(\ref{10}) will 
be expressed in terms of the coarse fields $B^a_{\mu}$ and one obtains 
the following recursion relation for the cubic coefficients
%\beea
%& c'_{\mu\nu\rho}(r_1,r_2)= \kappa H_{\mu\nu\rho}(r_1,r_2)+  \label{27} \\
%&  \sum_{n,n',n''} c_{\mu' \nu' \rho'}(n'-n,n''-n) Z_{\mu'\mu}(n'-r_1)
%  Z_{\nu'\nu}(n''-r_2) Z_{\rho'\rho}(n)  \nonumber 
%\eea
\bee
c'_{\mu\nu\rho}(r_1,r_2)= \kappa H_{\mu\nu\rho}(r_1,r_2)+    \label{27}
\ee
$$
  \sum_{n,n',n''} c_{\mu' \nu' \rho'}(n'-n,n''-n) Z_{\mu'\mu}(n'-r_1)
Z_{\nu'\nu}(n''-r_2) Z_{\rho'\rho}(n)
$$
where $H_{\mu\nu\rho}$ comes from expanding Eq.~(\ref{14}) up to
cubic order. The form of $H_{\mu\nu\rho}$ depends on the block 
transformation.
We discuss here neither the explicit form of this tensor nor the details
of solving Eq.~(\ref{27}) for the FP cubic couplings.
In Table \ref{tab:c} we give some of the largest elements of 
$c_{\mu\nu\rho}(r_1,r_2)$ for the block transformation of type~I
(for type~II we have not yet studied the cubic problem). 
Other related elements can be obtained by using the symmetry
properties of the cubic coupling tensor.

We shall return to the cubic couplings in paper~II when we
discuss the general  problem of the parametrization of the FP action in 
terms of loops.

\begin{table}
\begin{center}
\begin{tabular}{c c c}      \hline
  $r_1$     &    $r_2$   &    $c_{000}(r_1,r_2)$ \\   \hline
-2~~0~~0~~0   & -1~~0~~0~~0   &    -0.019        \\
-1~-1~~0~~0   & -1~~0~~0~~0   &    -0.012        \\
-1~-2~~0~~0   & -1~-1~~0~~0   &    -0.005        \\
-2~-1~~0~~0   & -1~-1~~0~~0   &    ~0.002        \\
\hline
  $r_1$     &    $r_2$   &    $c_{100}(r_1,r_2)$ \\   \hline
 0~-1~~0~~0   & ~0~-1~~0~~0   &    -0.598       \\
 0~-1~-1~~0   & ~0~~0~-1~~0   &    ~0.027       \\
 0~-1~~0~~0   & -1~-1~~0~~0   &    -0.027      \\
 0~-2~~0~~0   & ~0~-2~~0~~0   &    ~0.025      \\
\hline
  $r_1$     &    $r_2$   &    $c_{210}(r_1,r_2)$ \\   \hline
~0~-1~-1~~0   & ~0~-1~~0~~0   &    -0.062       \\
~0~-1~-1~-1   & ~0~-1~~0~-1   &    -0.007       \\
-1~~0~-1~~0   & ~0~-1~-1~~0   &    -0.004       \\
~0~-1~-1~~0   & ~0~-1~~0~-1   &    -0.004       \\
\hline
\end{tabular}
\end{center}
\caption{Some of the largest cubic couplings for the type~I RG transformation.}
\label{tab:c}
\end{table}

\section{Cut--off independence of physical predictions of the FP action
in 1--loop perturbation theory}

As discussed in Section~4, in leading order (tree level)
perturbation theory the FP action gives cut--off independent
physical predictions. Formal RG considerations suggest that 
the renormalized trajectory coincides with the FP action even at
1--loop level. That would imply that the predictions using the FP action
are free of lattice artifacts even in 1--loop perturbation
theory. Said differently: the predictions of the classical perfect
action (FP action) contain no O$\left( a^n \right)$ nor
O$\left( g^2 a^n \ln a, g^2 a^n\right)$ corrections for any $n\ge 1.$

This is a strong statement  which, unfortunately, we are not able
to support by a rigorous proof. 
We shall present the formal RG arguments for an infinite system
which show that the only effect of the perturbative $g^2$
corrections is to make the marginal operator $S^{FP}(U)$ weakly relevant,
i.e. the coupling $\beta=2 N/g^2$ begins to move, but the form
of the action on the renormalized trajectory remains
unchanged in this order:
\bee
\beta S^{FP}(U) \longrightarrow \beta' S^{FP}(V) \,,      \label{28}
\ee  
under a RG transformation, to 1--loop level. 
Here $\beta'=\beta -\Delta\beta$, where $\Delta\beta$ is fixed
by the first universal coefficient of the $\beta$--function.
(For a scale~2 block transformation $\Delta\beta=0.579$ in SU(3).)
An equation analogous to Eq.~(\ref{28}) occurred in ref.~\cite{WILSON}
where Wilson referred to formal RG arguments which were not, however, 
presented there.

Stronger arguments come from an explicit perturbative calculation with the 
FP action of the non--linear $\sigma$--model in $d=2$ \cite{AHNP}.
In ref.~\cite{AHNP} the mass gap $m(L)$ was calculated in a finite
periodic box of size $L$. 
%The main source of the error in this
%1--loop calculation is that the part of the FP action which is 
%quartic in the $\vec{\pi}$ fields is known to a final numerical precision
%only. 
%The very small cut--off effects  in $m(L)$ fall off exponentially 
%with $L$, and at $L=6$ they are 4 orders of magnitude smaller 
%than for the standard action\footnote{The leading cut--off dependence
%for the standard action is $\sim a^2/L^2$.}. 
The prediction of the standard action  is contaminated by power like
cut--off effects $(a^2/L^2)^n, n=1,2,...$. These type of cut--off
effects are eliminated when the FP action is used. More precisely,
in the actual calculation they are reduced to the level of the numerical
precision\footnote{The numerical errors are dominated by the errors
in the couplings of the quartic part of the action.}. For example, 
in the range of $L\sim 5~-~10$ the cut--off effect is reduced by 5
orders of magnitude relative to that of the standard action. On the
other hand, there exists a small non-power like cut-off effect, which
is seen at $L=2$, and goes to zero so rapidly that for $L>3$ it disappears
in the numerical errors. 
This behaviour can be associated with the small, but
finite, range of the FP action.

Let us present now the formal RG arguments. Consider a theory 
which has one marginal operator $R_1(U)$
and many irrelevant operators $R_n(U)$, $n=2,\dots$.
The action is
\begin{equation}
S = R_1(U) + c_2 R_2(U) + \dots
\label{ACT}
\end{equation}
and the Boltzmann factor is
$e^{-\beta S(U)}$, $\beta= 2N/ g^2$ for SU(N) gauge theories,
while $R_1(U)$ is the FP action $S^{FP}(U)$ in our earlier 
notations.
Under an RG transformation 
\begin{eqnarray}
S  & \longrightarrow & S'  = R_1(V) + c_2' R_2(V) + \dots
\nonumber \\
\beta & \longrightarrow & \beta^{\prime}.
\label{RGFLOW}
\end{eqnarray}
If the theory is asymptotically free the critical surface is at $g=0$.
A linearized scale
$s$ 
transformation at the fixed point is described by 
\begin{eqnarray}
g & \rightarrow & g \\
c_j & \rightarrow &\lambda_j c_j 
\end{eqnarray}
where $\lambda_j = s^{k_j} <1$ since the $R_j$'s are irrelevant.
As the fixed point is  at $g =0$,
the exponents $k_j$ are given by dimensional analysis.

Now we consider   general RG flows near the FP. For simplicity we
introduce the notation $c_1=g^2$.
\begin{eqnarray}
c_1' & = &c_1 + \alpha_{ij}^{(1)}c_i c_j  +O(c^3) \\
c_k' & = &\lambda_k c_k + \alpha_{ij}^{(k)}c_i c_j +O(c^3)  .
\end{eqnarray}
If we start with the action 
\begin{equation}
 S^{FP} = R_1,
\end{equation}
$c_1$ small, $c_2=c_3=...=0$, i.~e. the classical fixed point action,
then after one  RG step
\begin{eqnarray}
c_1' & = &c_1 + \alpha_{11}^{(1)}c_1^2  +O(c^3) \\
c_k' & = & \alpha_{11}^{(k)}c_1^2  +O(c^3) ,
\end{eqnarray}
or
\begin{equation}
{1\over g^2} S^{FP} \rightarrow {1\over{g^2 + \alpha_{11}^{(1)}g^4}}
(S^{FP} + \alpha_{11}^{(2)} g^4 R_2 +  \alpha_{11}^{(3)} g^4 R_3 + \dots) 
\end{equation}
Noticing that the prefactor can be written
\begin{equation}
{1\over{g^2 + \alpha_{11}^{(1)}g^4}} = {1\over {g^2}} - \alpha_{11}^{(1)}
+O(g^2)
\end{equation}
we see that to leading order in $g^2$ the effect of the RG transformation
is  only to shift the coupling
\begin{equation}
\beta S^{FP} \rightarrow\left(\beta - \alpha_{11}^{(1)}\right)
\left( S^{FP} + {{\alpha_{11}^{(2)} }\over {\beta^2}} R_2 + \dots
\right)\,,
\end{equation}
and that the action itself is changed only in order $g^4$.
The shift in $\beta$, $\alpha^{(1)}_{11}$, is equal to $\Delta\beta$
introduced earlier.

\section{Fixed point operators}

The action on the renormalized trajectory is perfect ---
it has only physical excitations. In a Green's function
only physical states come in as intermediate states independent
of the operator used. In spite of that Green's functions
(of fields, currents, etc.) show cut--off effects in general.
Consider, for example a free scalar field. As Eq.~(21) 
of ref.~\cite{HN} shows the perfect action on the lattice is 
described by a lattice field which is the integral of the continuum
field over a hypercube.
Therefore, the two-point function of the lattice field in the perfect
action is equal to the two-point function of the continuum field averaged
over the hypercubes around the the two lattice points in question. This
averaging brings in a trivial rotation symmetry breaking. The two-point
function of the field in the perfect action is not rotationally invariant,
although   only physical states propagate.

  A more interesting example is the static quark creation operator
represented conventionally by a Polyakov loop (or by one of the lines of an
elongated Wilson loop) The two-point function gives the quark-antiquark
potential. The potential measured with the Wilson action at distances
$a$ , $\sqrt{2}a$ ,$\sqrt{3}a$, $2a, \dots$
is non-smooth even at large beta values --- it reflects the hypercubic 
lattice structure rather than physics.
This is unfortunate since in numerical simulations these points lie in
the physically interesting region.
In addition, these are the points with the smallest statistical errors.
One expects that simple Polyakov loops constructed in terms of the 
fields of the perfect action behave better, since the cut--off effects
due to the distorted spectrum are eliminated. However, in order to get
rid of
the remaining cut--off effects we have to modify the operator itself.

The first step in this direction is to construct FP operators
along the lines by which the FP action was constructed. In Section~8.1
we use the example of the Polyakov loop to illustrate how to obtain 
the corresponding classical FP equation. In Section~8.2 we turn to the 
simple example of a scalar field theory in $d=2$ and construct the FP
field on the lattice. We show that the two--point function of the
FP field is free of any cut--off effects of order $(a/r)^n$. 
On the other hand, the FP field two--point function differs from the 
exact continuum propagator $-(2\pi)^{-1} \ln r +{\rm const}$
by terms which decrease exponentially with $r$.
In Section~8.3 we solve the FP equation for the Polyakov loop in 
quadratic approximation in the vector potentials, calculate 
the corresponding $q \bar{q}$ potential 
(which is $-(4\pi r)^{-1}$ in the continuum limit) 
and compare it with that of the standard and Symanzik improved actions.

\subsection{The FP Polyakov loop}

Consider the correlator 
$
\left. \langle L(\vec{n})L^{\dagger}(\vec{m})\rangle\right|_{S^{FP}},
$
where $L(\vec{n})$ is a Polyakov loop running in the temporal direction
at the spatial point $\vec{n}=(n_1,n_2,n_3)$ (with trace).
Perform a RG transformation as in Eq.~(\ref{4})
\bee
\int DU \rme^{-\beta \left( S^{FP}(U)+ T(U,V)\right)}
 L(U;\vec{n}) L^{\dagger}(U;\vec{m}) \,.                      \label{29}
\ee
The configuration $\{V\}$ on the coarse lattice is arbitrary;
it can be smooth or rough. In the $\beta\to\infty$ limit
(with $\{V\}$ fixed\footnote{Observe that this limit does not correspond 
to the standard perturbative expansion where the relevant $\{V\}$
configurations are smooth, $1-V \sim \beta^{-1/2}$.})
Eq.~(\ref{29}) is a saddle point problem.
One obtains
\bee 
\rme^{-\beta S^{FP}(V)} 
L^{\prime}(V;\vec{n}_B) L^{\prime\dagger}(V;\vec{m}_B)\,,   \label{30}
\ee
where
\bee
L^{\prime}(V;\vec{n}_B) \equiv L\left( U(V);\vec{n}=2\vec{n}_B \right)\,.
                                                                 \label{31}
\ee
Here $U=U(V)$ is the minimizing configuration of Eq.~(\ref{9})
and we assume that all the coordinates of the two points 
$\vec{n}$ and $\vec{m}$ are even, 
$\vec{n}=2\vec{n}_B$, $\vec{m}=2\vec{m}_B$.
The FP Polyakov loop operator is defined as
\bee
\lambda L^{FP}\left(V; \vec{n}_B={1\over 2}\vec{n}\right) =
 L^{FP}\left(U(V); \vec{n}\right)\,,                          \label{32}
\ee
for arbitrary configurations $\{ V \}$.
The eigenvalue $\lambda$ is expected to be 1 for the FP Polyakov loop.
Indeed, the formal limit of the Polyakov loop is
$\int d^4x {\rm ~current} \times {\rm vector~ potential}$ for a heavy
particle, and this is a standard, dimensionless part of the action.
This expectation is confirmed solving the FP equation Eq.~(\ref{32})
at the quadratic level (Section 8.3).

\subsection{The fixed point field in a free scalar theory}

We shall construct the FP field operator in a $d=2$ massless
scalar field theory and investigate its properties.
It is easy to handle this problem analytically and to demonstrate
the essential features of the FP operator.
Let us denote the scalar fields before and after the RG transformation
by $\pi_n$ and $\chi_{n_B}$, respectively. The RG transformation
is defined as
\bee
\exp\left\{ -{1\over 2}\sum_{n_B,r_B} \rho'(r_B)
 \chi_{n_B}\chi_{n_B+r_B}\right\} =
                                                               \label{33}
\ee
$$
C \int D\pi
\exp\left\{-{1\over 2}\sum_{n,r} \rho(r)\pi_n \pi_{n+r}
- 2 \kappa \sum_{n_B} \left( \chi_{n_B}-{1\over 4}
\sum_{n\in n_B} \pi_n \right)^2 \right\} \,, 
$$
where $\kappa$ is a free parameter. The corresponding FP action
has been constructed by Wilson and Bell \cite{WB} a long time ago.
This free field problem is also relevant for the FP action of the non--linear
$\sigma$--model \cite{HN}. We shall follow the notations of ref.~\cite{HN}.
The FP solution of Eq.~(\ref{33}) in Fourier space has the form
\bee
{1\over \rho^{FP}(q)}= \sum_{l=-\infty}^{\infty}
{1\over (q+2\pi l)^2}
\prod_{i=0}^1 {\sin^2({q \over 2}+\pi l)_i \over
 ({q \over 2}+\pi l)_i^2}
+{1\over 3\kappa}\,,                                         \label{34}
\ee
where the summation is over the integer vector $l=(l_0,l_1)$ and
$q_i \in (0,2\pi)$.
A Gaussian integral is equivalent to a minimization problem.
The FP equation for Eq.~(\ref{33}) can also be written as
\bee
{1\over 2}\sum_{n_B,r_B} \rho^{FP}(r_B) \chi_{n_B}\chi_{n_B+r_B}\ =
                                                             \label{35}
\ee
$$
\min_{\{\pi\}} \left\{
{1\over 2}\sum_{n,r} \rho^{FP}(r)\pi_n \pi_{n+r}
+2 \kappa \sum_{n_B} \left( \chi_{n_B}-{1\over 4}
\sum_{n\in n_B} \pi_n \right)^2 \right\} \,. 
$$
Let us denote the minimizing configuration by $\bar{\pi}_n(\chi)$.
One can repeat the minimization procedure by going to even finer
lattices, i.e. making a multigrid minimization, in principle to
arbitrary depth. Then the minimizing field on the finest lattice
at some point $x_B$ (measured in units of the coarse lattice)
is given by a function $\Phi(\chi,x_B)$ of the coarse field.
Clearly, it gives the solution to the fixed point problem:
\bee
\lambda \Phi^{FP}\left( \chi;x_B \right)=
\Phi^{FP}(\bar{\pi}(\chi);2 x_B+ {1\over 2} )\,,            \label{36}
\ee
where $\lambda=1$ as expected for a  dimensionless  scalar field in $d=2$.
The FP field operator $\Phi^{FP}(\pi;x)$ is a linear combination of the
fields $\pi_{n'}$.
The procedure of finding the FP field is simplified by the
 observation that $k$ RG steps
are equivalent to a single blocking step with a scale factor of
$L=2^k$ with somewhat modified parameters \cite{WB}:
\bee
\exp\left\{ -{1\over 2}\sum_{n_B,r_B} \rho^{FP}(r_B)
 \chi_{n_B}\chi_{n_B+r_B}\right\} =                           \label{37}
\ee
$$
C \int D\pi
\exp\left\{-{1\over 2}\sum_{n,r} \rho^{FP}(r)\pi_n \pi_{n+r}
- {a_k\over 2} \sum_{n_B} \left( \chi_{n_B}-b_k
\sum_{n\in n_B} \pi_n \right)^2 \right\} \,,  
$$
where the summation $n\in n_B$ goes over the  $L^2=4^k$ fine points 
of the block labeled by $n_B$ and 
\bee 
 a_k={3\kappa \over 1-1/L^2},~~~~
 b_k={1\over L^2},~~~ L=2^k \,.                               \label{38}
\ee
The minimizing configuration $\bar{\pi}_n(\chi)$ of 
the quadratic form in the exponent of Eq.~(\ref{37}) gives then the
FP operator in the $k\to\infty$ limit
\bee
\Phi^{FP}(\chi,n_B)= \lim_{k\to\infty}\bar{\pi}_{n}(\chi)\,.
                                                            \label{39}
\ee
Here we take\footnote{ 
For the given block transformation the sites in the block $n_B$
are given by $n=L n_B +m$ where $m_i=0,1,\ldots,L-1$ for the components
$i=0,1$ and the coarse field is coupled to the average over this region. 
Hence the choice $m_i=L/2$ represents the site sitting in the middle
of the averaging region.}  $n_i=L (n_{Bi}+{1\over 2})$. 
In Fourier space we have \cite{WB}
\bee
\bar{\pi}(q)={1\over L^2}u^*_L(q)
{\rho^{FP}(Lq) \over \rho^{FP}(q)} \chi(Lq)                \label{40}
\ee
where
\bee
u_L(q)=\prod_{j=0}^1 {1-\rme^{i q_j L} \over 1-\rme^{i q_j} } \,.
                                                            \label{41}
\ee
Taking the $k\to\infty$ limit one obtains
\bee
\Phi^{FP}(n_B) = \sum_{n'_B} C(n_B - n'_B) \chi_{n'_B} \,,
                                                           \label{43}
\ee
or in Fourier space
\bee
\Phi^{FP}(q_B)= C(q_B) \chi(q_B)\,,                       \label{44}
\ee
where
\bee
C(q_B)=\rho^{FP}(q_B) \sum_{l=-\infty}^{\infty}
{1\over (q_B + 2\pi l)^2} 
\prod_{i=0}^1 { \sin\left( {q_B \over 2} +\pi l \right)_i
\over \left( {q_B \over 2} +\pi l \right)_i } \,.
                                                         \label{45}
\ee

Let us demonstrate now that the propagator of the FP field $\Phi^{FP}$
is free of those cut--off effects which go to zero  as an
inverse power of the distance. We show first that the propagator
of the $\chi$ field itself has such cut--off effects in spite
of the fact that the spectrum is exact. Using Eq.~(\ref{34})
and introducing $k=q+2\pi l$, we get
\bee
\langle \chi_n \chi_0 \rangle =
\int_{-\infty}^{\infty} {d^2k\over (2\pi)^2}
{\rme^{ikn}\over k^2} \prod_j {\sin^2{k_j\over 2} \over
(k_j/2)^2}, ~~~ n\ne 0\,.                           \label{46}
\ee
Consider $|n| \gg 1$ (i.e. low momenta) and expand in $k$:
\bee
\langle \chi_n \chi_0 \rangle =
\int_{-\infty}^{\infty} {d^2k\over (2\pi)^2}
{\rme^{ikn}\over k^2} 
\left[ 1 - {1\over 12} k^2 + {1\over 360}(k^2)^2
-{1\over 1440}(k_0^4 + k_1^4) + \ldots \right]             \label{47}
\ee
The $(k^2)^s$, $s\ge 1$ terms in the bracket do not give long range
contributions. On the other hand, the rotationally non--invariant term
$k_0^4+k_1^4$ (since it does not cancel the $1/k^2$ singularity)
leads to  cut--off dependent long range corrections
\bee
-{1\over 240\pi}{1\over (n^2)^2} + 
{1\over 30\pi}{n_0^2 n_1^2 \over (n^2)^4}\,.              \label{47a}
\ee
Consider the propagator of the FP field in Eq.~(\ref{44})
\bee
D^{FP}(q)={ \left| C(q) \right|^2 \over \rho^{FP}(q)}\,.  \label{48}
\ee
For small $q$ we have
\bee
{1\over \rho^{FP}(q)} \sim {1\over q^2}
\prod_{i=0}^1 {\sin^2{q_i\over 2} \over (q_i/2)^2}
+ {\rm reg.~terms} \,,                                   \label{49}
\ee
$$
\left| C(q) \right|^2 \sim
\left( \rho^{FP}(q) \right)^2
\left[ {1\over q^2}
\prod_{i=0}^1 {\sin{q_i\over 2} \over (q_i/2)}
+ {\rm reg.~terms} \right]^2\,,     
$$
which shows that
\bee
D^{FP}(q) \sim {1\over q^2} \left( 1+ q^2 R(q) \right) \,,    \label{50}
\ee
where $R(q)$ is regular at $q=0$. In $D^{FP}(q)$ the singular
$1/q^2 \cdot \sum_i (q_i)^k$, $k>1$ terms cancel, therefore
no cut--off dependent corrections enter which vanish like an inverse
power of $n$.

The FP propagator has exponentially decaying cut--off dependent
corrections, however. The following considerations not only illustrate
this statement but give a general argument why the power corrections
are canceled.

Collect the terms quadratic in $\pi$ in Eq.~(\ref{37}) and introduce the
notation
\bee
\sum_{n, n'} Q(n,n')\pi_n \pi_{n'} =
\sum_{n,r} \rho^{FP}(r)\pi_n \pi_{n+r} +
a_k b_k^2 \sum_{n_B}\left( \sum_{n\in n_B} \pi_n \right)^2 \,.
                                                              \label{51}
\ee
The exponent in Eq.~(\ref{37}) then has the form 
\bee
-{1\over 2} \left[ \sum_{n,n'}  Q(n,n')
\left( \pi_n - \bar{\pi}_n(\chi) \right)
\left( \pi_{n'} - \bar{\pi}_{n'}(\chi) \right) +
\sum_{n_B, r_B} \rho^{FP}(r_B) \chi_{n_B} \chi_{n_B+r_B} \right] \,.
                                                               \label{52}
\ee
% where $\bar{\pi}_n(\chi)$ is the minimizing configuration, as before
Let $n=2^k(n_B+{1\over 2})$ and  $n'=2^k(n_B'+{1\over 2})$.
For given $n_B$, $n_B'$ and $k\to\infty$
\bee
\Delta^{\rm cont}(n-n')={1\over Z}\int D\pi \pi_n \pi_{n'}
\exp\left\{-{1\over 2} \sum_{m,r} \rho^{FP}(r) \pi_m \pi_{m+r}\right\}
                                                             \label{53}
\ee
is the continuum propagator $-(2\pi)^{-1} \ln |n| + {\rm const}$.
Insert $1$ in the path integral in Eq.~(\ref{53}) using
\bee
1 \sim \int D\chi \exp\left\{
-{1\over 2} a_k \sum_{m_B} \left( \chi_{m_B}- b_k
\sum_{m\in m_B} \pi_m \right)^2 \right\} \,.                 \label{54}
\ee
Using Eq.~(\ref{52}) and introducing the new integration variables
\bee
\psi_n=\pi_n-\bar{\pi}_n(\chi) \,,                           \label{55}
\ee
we get
\bee
\Delta^{\rm cont}(n-n')=
\left.
\langle\Phi^{FP}(\chi,n_B)\Phi^{FP}(\chi,{n_B'})\rangle\right|_{\rho^{FP}}
+ \left. \langle \psi_n \psi_{n'} \rangle \right|_{Q} \,.      \label{56}
\ee
Note that $\Delta^{\rm cont}(n-n')=\Delta^{\rm cont}(n_B-n_B')+{\rm const}$.
The difference between the continuum and FP propagators is the inverse 
of $Q$ defined in Eq.~(\ref{51}). The tensor $Q(n,n')$ connects fields
over a distance $\sim L a_{\rm fine}= a_{\rm coarse}$ so it is expected
to produce correlations at least over such distances. On the other hand,
comparing eqs.~(\ref{51}) and (\ref{37}), 
$\left. \langle \psi_n \psi_{n'} \rangle \right|_Q$ can be interpreted as
propagation in the presence of the external fields $\chi_{m_B}$
which are put now to zero at every coarse lattice point ${m_B}$.
This forces the block averages $\sum_{m\in m_B}\pi_m$ to fluctuate
around zero. Such an external field problem is expected to produce
short ranged, exponentially decaying correlations\footnote{
Actually, this is the basic assumptions of the theory of RG
transformations.}.

% OTHERWISE 
% 
% It is worth to explain these properties  for the case when the 
% continuum limit is taken for the fine lattice, using a somewhat
% intuitive reasoning.
%  Replace the coordinate $n$ on the fine lattice
% by $x_i=n_i/L - 1/2$ which becomes continuous for $k\to\infty$.
% The field $\chi_{n_B}$ on the coarse lattice represents an average
% of the fields $\pi(x)$ in the continuum over the unit 
% square $|x_i - n_{Bi}|< {1\over 2}$.
% The correlator of the $\pi(x)$ fields is given by the 
% continuum propagator. However, the field $\chi_{n_B}$, being related to
% an average over a square, has multipole moments and consequently
% its correlator differs from the pointlike one by terms like 
% $r^{-4} \cos (4\varphi)$ where $\varphi$ is the polar angle of
% $\vec{x}$ (cf. Eq.~(\ref{47})). 
% The FP field $\Phi(x,\chi)$, on the other hand, 
% represents the value around which the fine field $\pi(x)$ 
% fluctuates in the corresponding path integral,
% i.e.  the minimizing field $\bar{\pi}(x,\chi)$.
% The wavelength of the typical fluctuations $\psi(x)$ 
% is effectively bounded by the coarse lattice size --- 
% the average field over a square is forced to fluctuate around the
% given coarse field, and this fact provides an effective mass
% to long wavelength modes.

\subsection{The FP Polyakov loop at the quadratic level and
the $q\bar{q}$ potential}

We solve  now the FP equation for the Polyakov loop Eq.~(\ref{32}) 
by iteration in quadratic approximation in the vector potentials. 
At the quadratic level we have
\bee
L(U,n;k)=N-{1\over 4}\sum_{n', n''} 
W_{\mu \nu}(\vec{n}'-\vec{n},\vec{n}''-\vec{n},n_0'-n_0'';k)
A_{\mu}^a(n') A_{\nu}^a(n'')\,,                        \label{57}
\ee
where $k$ keeps track on the number of iterations performed,
 $n=(n_0,\vec{n})$ , etc. and a summation over the Lorentz and colour 
indices is understood. In Eq.~(\ref{57}) time translation invariance 
is used. $W$ is the unknown function to be determined.
Using the relation Eq.~(\ref{25}) between the coarse vector potential $B$
 and the minimizing $A$ one obtains the following recursion relation for $W$:
\bee
W_{\rho\sigma}(\vec{n}_B',\vec{n}_B'',n_{B0}'-n_{B0}'';k+1)=
                                                           \label{58}
\ee
$$
\sum_{n',n''} W_{\mu\nu}(\vec{n}',\vec{n}'',n_{0}'-n_{0}'';k)
 Z_{\mu\rho}(n'-2n_B')Z_{\nu\sigma}(n''-2n_B'')\,.
$$
The starting value (corresponding to a Polyakov loop at $\vec{n}=0$) is
\bee
W_{\mu\nu}(\vec{n}',\vec{n}'',n_{0}'-n_{0}'';0)=
\left\{ \begin{array}{ll}
            1  & \mbox{if $\mu=\nu=0$, $\vec{n}'=\vec{n}''=0$} \\
            0  & \mbox{otherwise.}
        \end{array}
\right.
                                                         \label{59}
\ee

Due to the fact that $W(.,k=0)$ is independent of 
$n_0'-n_0''$, $W(.,k)$ is time independent, too. 
In addition, $W(.,k=0)$ is factorized in $\vec{n}'$ and 
$\vec{n}''$, from which the same follows for arbitrary $k$. 
This is a great simplification. Using the reflection properties
of $Z$ (Appendix B) one obtains
\bee
\widehat{Z}_{0i}(\vec{n})=0,~~~i=1,2,3\,,                       \label{60}
\ee
where
\bee
\widehat{Z}_{\mu\nu}(\vec{n})=\sum_{n_0} Z_{\mu\nu}(\vec{n},n_0)\,. \label{61}
\ee

Eq.~(\ref{60}) implies that only the $\mu=\nu=0$ component of $W$ 
is non--zero for any $k$. One finds:
\bee
L^{FP}(U,0)=N-{1\over 4}
\left( \sum_{\vec{n}} w(\vec{n}) \sum_{n_0} A_0^a(n_0,\vec{n})\right)^2
                                                               \label{62}
\ee
plus terms higher order in $A$, where $w(\vec{n})$ is the $\lambda=1$ 
eigenvector of the following eigenvalue equation:
\bee
\sum_{\vec{n}} \widehat{Z}_{00}(\vec{n}-2\vec{n}_B) w(\vec{n}) =
\lambda w(\vec{n}_B) \,.                                      \label{63}
\ee

One can solve the eigenvalue equation by iteration starting with 
$w(\vec{n})=\delta_{\vec{n}0}$.
The iteration will automatically project out the eigenvector with 
the largest eigenvalue $\lambda=1$. Given $Z$, this is a trivial problem.
The error in $w(\vec{n})$ is dominated by the error in the connecting
function $Z$. 
For the case of type~I RG transformation the result is shown in
Table~5. The other elements are smaller than $10^{-5}$. 
The eigenvector $w(\vec{n})$ in case of type~II RG transformation looks 
rather similar \cite{BLATTER}. 
It follows from Eq.~(\ref{63}) and Eq.~(B2) 
in Appendix B that $w(\vec{n})$ is reflection symmetric under
 $n_i \to -n_i$, $i=1,2,3$. 
The eigenvalue $\lambda$ is unity in the quadratic approximation,
as expected.

% TABLE 6

\begin{table}
\begin{center}
\begin{tabular} {c c | c  c}       \hline
  $\vec{n}$     &    $w(\vec{n})$  & $\vec{n}$ &  $w(\vec{n})$ \\   \hline
0~0~0 &  $0.8951$ & 0~1~2 &     $0.1312\times 10^{-3}$       \\
0~1~1 &  $0.5626\times 10^{-2}$ & 0~2~2 &     $0.3763\times 10^{-4} $     \\
1~1~1 &  $0.1729\times 10^{-2}$ & 0~1~3 &     $0.2709\times 10^{-4}$       \\
0~0~1 &  $0.1552\times 10^{-2}$ & 1~2~2 &     $0.1806\times 10^{-4}$       \\
0~0~2 &  $0.9364\times 10^{-3}$ & 1~1~3 &     $0.1124\times 10^{-4}$       \\
1~1~2 &  $0.1425\times 10^{-3}$ & 0~2~3 &     $0.3954\times 10^{-5}$       \\
\hline
\end{tabular}
\end{center}
\caption{The weights $w(\vec{n})$ for the FP Polyakov loop.}
\end{table}

Let us consider now the correlation function of two FP Polyakov loops
\bee
\langle L^{FP}(U; \vec{n}) L^{FP}(U; 0) \rangle       \label{64}
\ee
expanding $L^{FP}$ up to quadratic order in the vector potentials. 
The connected part is proportional to the square of the lowest order 
quark--antiquark potential \cite{REISZ}. 
Using Eq.~(\ref{62}) we get 
\bee
V(\vec{r})=\int_0^{2\pi}{d^3q\over (2\pi)^3}
\rme^{i\vec{q}\vec{r}} w(\vec{q})w(-\vec{q})D_{00}(q_0=0,\vec{q})\,, \label{65}
\ee
where $D$ is the propagator of the FP quadratic action 
(Section 4 and Appendix A) and
\bee
w(\vec{q})=\sum_{\vec{n}} \rme^{-i\vec{q}\vec{n}} w(\vec{n})        \label{66}
\ee
is the Fourier transform of the FP Polyakov loop. 
In the continuum limit we should get $V(\vec{r})=-(4\pi r)^{-1}$.

We demonstrate now that $V(\vec{n})$ in Eq.~(\ref{65}) has no cut--off effects
which for large $|\vec{n}|$ decrease  as an inverse power of 
$|\vec{n}|$. 
Said differently, the potential derived from the FP Polyakov loop 
is a solution of the (on--mass--shell) Symanzik program through
 all orders in $a^2$.
The argument follows the same lines as in a free scalar theory, 
eqs.~(\ref{48}-\ref{50}).
Separate first the singular part of $D_{00}(q_0=0,\vec{q})$ at small 
$\vec{q}$. As Eq.~(A7) in Appendix A shows, the singular contribution to 
$D^{(n)}$ after $n$ iterations comes from the $l=0$ term
\bee
D_{00}^{(n)}(q_0=0,\vec{q})= \left| 
\Omega_{00}^{(n)} \left(2^{-n}\vec{q}\right) \right|^2 {1\over \vec{q}{}^2}
+ {\rm regular~terms},                                      \label{67}
\ee
where we used Eq.~(\ref{20}) and the fact that only the $00$ 
component of $\Omega$ survives at $q_0=0$ as eqs.~(\ref{16a},\ref{16b}),
and (A8) show. The presence of the non--rotationally invariant residue
in Eq.~(\ref{67}) is responsible for the remaining cut--off effects
decaying as a power of $|\vec{n}|$ if we use simple Polyakov loops
($w(\vec{q}) = 1$). This dangerous residue is cancelled by
$w(\vec{q}) w(-\vec{q})$ in the FP loop correlators, however.

Using Eq.~(\ref{63}) we get the recursion relation for the eigenvector 
with $\lambda=1$ (the argument $q_0=0$ is not indicated explicitly 
in the following equations):
\bee
w^{(n+1)}(\vec{q})={1\over 8} \sum_{l_i=0}^1
 Z_{00}^{(n)}\left( {\vec{q}+2\pi \vec{l}\over 2}\right)
 w^{(n)}\left( {\vec{q}+2\pi \vec{l}\over 2}\right)\,,           \label{68}
\ee
where $Z_{00}$ is obtained from Eq.~(\ref{26})
\bee
 Z_{00}^{(n)}\left( {\vec{q}+2\pi \vec{l}\over 2}\right)=
D_{00}^{(n)}\left( {\vec{q}+2\pi \vec{l}\over 2}\right)
\omega_{00}^*\left( {\vec{q}+2\pi \vec{l}\over 2}\right)
\left[ D_{00}^{(n+1)}(\vec{q})\right]^{-1} \,.                   \label{69}
\ee
Iterating Eq.~(\ref{68}) leads to
\bee
w^{(n+1)}(\vec{q})=
{1\over 4^n}\left[ D_{00}^{(n+1)}(\vec{q}) \right]^{-1}
% {4^{-n}\over D_{00}^{(n+1)}(\vec{q})}
\sum_{l_i=0}^{2^n-1} 
\Omega_{00}^{(n)*}\left( {\vec{q}+2\pi \vec{l}\over 2^n}\right)
D_{00}^{(0)}\left( {\vec{q}+2\pi \vec{l}\over 2^n}\right) \,.   \label{70}
\ee
For large $n$ eqs.~(\ref{70}),(\ref{67}) and (\ref{20}) give
\bee
w^{(n+1)}(\vec{q}) = \vec{q}{}^2
\left( {1\over \left| \Omega_{00}^{(n+1)}\left( 2^{-n}\vec{q}\right)
 \right|^2} + \vec{q}{}^2 R(q)\right) 
\sum_{\vec{l}=0}^{2^n-1}\Omega_{00}^{(n)*}\left( {\vec{q}+2\pi \vec{l}\over
2^n}\right) {1\over (\vec{q}+2\pi \vec{l})^2} \,.               \label{71}
\ee
Here and below we denote by $R(q)$ a generic term regular at $q=0$.
Only the $\vec{l}=0$ term gives relevant contribution
\bee
w^{(n+1)}(\vec{q}) =
\left( {1\over \left| \Omega_{00}^{(n+1)}\left( 2^{-n}\vec{q}\right)
 \right|^2} + \vec{q}{}^2 R(q) \right) 
\left(\Omega^{(n)*}\left( {\vec{q}\over 2^n}\right) +
 \vec{q}{}^2 R(q)\right) \,.                      \label{72}
\ee
Combining eqs.~(\ref{65}), (\ref{67}) and (\ref{72}) we get
\bee
V(\vec{r}) = \int_0^{2\pi}{d^3q\over (2\pi)^3}
\rme^{i\vec{q}\vec{r}} {1\over \vec{q}{}^2} \left( 1+ \vec{q}{}^2 R(q)
\right) \,,                                          \label{73}
\ee
which we wanted to show.

The considerations above say nothing about the exponentially decaying cut--off effects. 
For a graphical illustration in Figs.~\ref{fig:rvrperfect},
\ref{fig:rvrsyman} and \ref{fig:rvrwilson} we plot the
perturbative potential $rV(r)$ versus $r$ for the FP, Symanzik and 
Wilson actions.

\section{Outlook}

The most relevant task for future progress is to construct the FP
action for full QCD including fermions. The free fermion case and
the anomaly problem in the Schwinger model have been considered in
this context \cite{BWUW}. The QCD FP action is quadratic in the
fermion fields, which is an important simplification and should make 
this problem feasible.

\section{Acknowledgements}
U. Wiese participated in the early stages of this project. We
are indebted to M.~Blatter, R.~Burkhalter, P.~Kunszt
and P.~Weisz for valuable
discussions.
We would like to thank  T. Barker,
M. Horanyi and the Colorado high energy experimental
group for allowing us to use their work stations. 
This work was supported by the U.S. Department of 
Energy and by the National Science Foundation and by the Swiss National
Science Foundation.

\appendix

\section{Appendix}

In this Appendix we discuss the steps of solving Eq.~(\ref{17}) for
the FP $\rho^{FP}_{\mu\nu}$.

In the first term on the r.h.s. of Eq.~(\ref{17}) we write
$$
k={1\over 2}k_B+\pi l,~~~~{1\over V}\sum_k =
{1\over V_B}\sum_{k_B} {1\over 16} \sum_l \,,    \label{A1}
\eqno{(A1)}
$$
where $l_{\mu}=0,1$. Taking the derivative of the r.h.s. with
respect to $A_{\nu}(-q)$ for $q= {q_B \over 2}+\pi l$ 
one obtains
$$
A_{\nu}\left( {q_B \over 2}+\pi l \right) +
\kappa \left[ D \left( {q_B \over 2}+\pi l \right)
\omega^{\dagger}\left({q_B \over 2}+\pi l\right) 
\right]_{\nu\mu} 
\left(\Gamma_{\mu}(q_B)-B_{\mu}(q_B)\right)=0 .       \label{A2}
\eqno{(A2)}
$$
It is assumed that some gauge fixing term is introduced and so
$\rho$ has an inverse which is denoted here by $D$.
First one solves Eq.~(A2) for $\Gamma_{\mu}$. After multiplying with 
$\omega_{\gamma\nu}\left({q_B \over 2}+\pi l\right)$
and summing over $\nu$ and $l$ we get
$$
\Gamma_{\gamma}(q_B)=
\left[ {\cal M}(q_B) +{1\over \kappa} \right]_{\gamma \mu}^{-1}
  {\cal M}_{\mu\rho}(q_B) B_{\rho}(q_B)\,,                \label{A3}
\eqno{(A3)}
$$
where
$$
{\cal M}_{\mu\nu}(q_B)={1\over 16} \sum_l
\left[ \omega\left({q_B \over 2}+\pi l\right)
      D\left( {q_B \over 2}+\pi l\right)
      \omega^{\dagger}\left({q_B \over 2}+\pi l\right)
\right]_{\mu\nu}\,.                                  \label{A4}
\eqno{(A4)}
$$
After substituting Eq.~(A3) into Eq.~(A2), the minimizing
$A_{\mu}$ is obtained as a function of the coarse field $B_{\nu}$
$$
A_{\mu}\left( {q_B \over 2}+\pi l\right) =
\left[ D\left({q_B \over 2}+\pi l\right)
     \omega^{\dagger}\left({q_B \over 2}+\pi l\right)
\right]_{\mu\rho}
\left( {1\over \kappa} + {\cal M}(q_B) \right)_{\rho\nu}^{-1}
B_{\nu}(q_B)\,.                                        \label{A5}
\eqno{(A5)}
$$
We can write now Eq.~(A5) into Eq.~(\ref{17}) and find after
some algebra
$$
D'_{\mu\nu}(q_B)={\cal M}_{\mu\nu}(q_B) + {1\over \kappa} \delta_{\mu\nu}
                                                       \label{A6}
\eqno{(A6)}
$$
which is the result quoted in Eq.~(\ref{18}).
% where $D$ and $D'$
%denote the inverse of $\rho$ and $\rho'$, respectively.

In iterating Eq.~(\ref{18}) let us start with some $D_{\mu\nu}^{(0)}$.
We might take the form given in Eq.~(\ref{20}). After $n$ RG
steps we get
$$
D_{\mu\nu}^{(n)}(q)={1\over 4^n} \sum_{l=0}^{2^n-1}
\left[
  \Omega\left({q+2\pi l\over 2^n}\right)
   D^{(0)}\left({q+2\pi l\over 2^n}\right)
  \Omega^{\dagger}\left({q+2\pi l\over 2^n}\right)
\right]_{\mu\nu} +{1\over \kappa}\sum_{j=0}^{n-1}
  Q_{\mu\nu}^{(j)}\,,                                         \label{A7}
\eqno{(A7)}
$$
where
$$
     \Omega_{\mu\nu}^{(j)}(2^{-j}k) = {1\over 2^j}
     \left( \omega(2^{-1}k) \omega(2^{-2}k) \ldots \omega(2^{-j}k)
     \right)_{\mu\nu}
$$
$$
     \Omega_{\mu\nu}^{(1)}(2^{-1}k)={1\over 2} \omega_{\mu\nu}(2^{-1}k)\,,
     \eqno{(A8)}                                             \label{A8}
$$
and
$$
       Q_{\mu\nu}^{(j)}(q)  =  {1\over 4^j}
         \sum_{l=0}^{2^j-1}
         \left[
            \Omega^{(j)}(2^{-j}(q+2\pi l))\Omega^{(j)\dagger}
              (2^{-j}(q+2\pi l))
         \right]_{\mu\nu} 
$$
$$
      Q_{\mu\nu}^{(0)}(q) = \delta_{\mu\nu} \,.
  \eqno{(A9)}                                              \label{A9}
$$

One can prove eqs.~(A7--A9) by induction.
We introduced the following shorthand notations above
$$
\sum_{l=0}^{2^n-1}= \sum_{l_0=0}^{2^n-1} \ldots \sum_{l_3=0}^{2^n-1}\,.
                                                              \label{A10}
\eqno{(A10)}
$$
The tensor $\Omega_{\mu\nu}^{(j)}(k)$ in Eq.~(A8) can be written in a form
which is well suited for a numerical procedure. For both types of 
RG transformations $\omega_{\mu\nu}(k)$ in eqs.~(\ref{16a},\ref{16b})
can be written as
$$
\omega_{\mu\nu}(k)={ \widehat{2k_{\mu}} \over \widehat{k}_{\mu}} 
a^{(1)}(k) \delta_{\mu\nu} + 
{ \widehat{2k_{\mu}} \over \widehat{k}_{\nu}} b_{\nu}^{(1)}(k) \,.  
                                                             \label{A11}
\eqno{(A11)}
$$
One obtains then
$$
\Omega_{\mu\nu}^{(j)}(2^{-j}k)=
2^{-j} {\widehat{k}_{\mu} \over \widehat{2^{-j}k}_{\mu}} a^{(j)}(2^{-j}k) +
2^{-j} {\widehat{k}_{\mu} \over \widehat{2^{-j}k}_{\nu}} 
b^{(j)}_{\nu}(2^{-j}k)\,,
                                                                 \label{A12}
\eqno{(A12)}
$$
where $a^{(j)}$ and $b_{\nu}^{(j)}$ satisfy the simple recursive relation
$$
       a^{(j)}(2^{-j}k) =  a^{(1)}(2^{-1}k) a^{(j-1)}(2^{-j}k)\,,
$$
$$
   b_{\nu}^{(j)}(2^{-j}k)=b_{\nu}^{(1)}(2^{-1}k) a^{(j-1)}(2^{-j}k)+
   b_{\nu}^{(j-1)}(2^{-j}k) \,.
\eqno{(A13)}                                        \label{A13}
$$
The equations above allow a sufficiently precise numerical determination
of the fixed point propagator $D_{\mu\nu}^{FP}(q)$.

Let us now discuss briefly the steps of removing the gauge fixing.
The brute force method is to take the $\alpha\to\infty$
limit of $\rho_{\mu\nu}^{FP}=\left( D^{FP} \right)_{\mu\nu}^{-1}$ numerically.
More elegant is to observe that the longitudinal part of the propagator
remains longitudinal after RG transformations:
$$
D_{\mu\nu}^{(0)} = D^{(0)}(q) \delta_{\mu\nu} 
+\alpha \widehat{q}_{\mu} \widehat{q}_{\nu}^* f^{(0)}(q)
\longrightarrow
D_{\mu\nu}^{(n)} = G_{\mu\nu}^{(n)}(q)
+\alpha \widehat{q}_{\mu} \widehat{q}_{\nu}^* f_{\nu}^{(n)}(q)\,.
\eqno{(A14)}                                               \label{A14}
$$
where $G_{\mu\nu}^{(n)}(q)$ is the part evolved from
$D^{(0)}(q) \delta_{\mu\nu}$. Eq.~(A14) can be shown easily
by using the gauge symmetry relation satisfied by the tensor
$\Omega^{(j)}$
$$
\Omega^{(j)}_{\mu\rho}(k) \widehat{k}_{\rho} =
 2^{-j} \widehat{2^j k}_{\mu} \,.                   \label{A15}
\eqno{(A15)}
$$
Using Eq.~(A14) one obtains for large $\alpha$
$$
\left( D^{(n)}(q) \right)_{\mu\nu}^{-1} =
\rho_{\mu\nu}^{(n)}(q) + {1\over \alpha} t_{\mu\nu}(q)
+{\rm O}\left( {1 \over \alpha^2} \right) \,,        \label{A16}
\eqno{(A16)}
$$
where $\rho_{\mu\nu}^{(n)}(q)$ is the gauge invariant part of the quadratic
blocked action we are looking for. Introducing the projector
$$
T_{\mu\nu}(q)=\delta_{\mu\nu} - 
{ \widehat{q}_{\mu} \widehat{q}_{\nu}^* \over
 ( \widehat{q} \widehat{q}^*) }                             \label{A17}
\eqno{(A17)}
$$
it is easy to show that
$$
\rho_{\mu\nu}(q) = 
\left( T(q) G(q; \eta)^{-1} T(q) \right)_{\mu\nu} \,,      \label{A18}
\eqno{(A18)}
$$
where
$$
G(q; \eta)_{\mu\nu} = \left( T(q) G(q) T(q) \right)_{\mu\nu}
+ \eta  { \widehat{q}_{\mu} \widehat{q}_{\nu}^* \over
 ( \widehat{q} \widehat{q}^*) }  \,.                         \label{A19}
\eqno{(A19)}
$$
Here $\eta$ is an arbitrary non--zero parameter and $\rho$ in
Eq.~(A18) is independent of $\eta$.

\section{Appendix}

Using the definitions Eq.~(\ref{12}) and Eq.~(\ref{25})
for $\rho_{\mu\nu}(r)$ and $Z_{\mu\nu}(r)$, respectively, it is easy to derive
the following transformation properties under reflection of the 
coordinates.

Consider the index combinations $\mu\nu=00$ and $10$.
The other cases can be obtained by permuting the coordinates.
We get ($r=(r_0,r_1,r_2,r_3)$)
$$
\rho_{00}(r_0,r_1,r_2,r_3) =\rho_{00}(\pm r_0,\pm r_1,\pm r_2,\pm r_3)\,,
$$
$$
\rho_{10}(r_0,r_1,r_2,r_3) =  -\rho_{10}(-r_0+1,r_1,\pm r_2,\pm r_3)\,,
\eqno{(B1)}                                             \label{B1} 
$$
$$
\rho_{10}(r_0,r_1,r_2,r_3) =  -\rho_{10}(r_0,-r_1-1,\pm r_2,\pm r_3)\,,
$$

$$
Z_{00}(r_0,r_1,r_2,r_3)  =  Z_{00}(r_0,\pm r_1,\pm r_2,\pm r_3)\,,
$$
$$
Z_{00}(r_0,r_1,r_2,r_3)  =  Z_{00}(-r_0+1,\pm r_1,\pm r_2,\pm r_3)\,,
$$
$$
Z_{10}(r_0,r_1,r_2,r_3)  =  -Z_{10}(-r_0+2,r_1,\pm r_2,\pm r_3)\,,
\eqno{(B2)}                                               \label{B2} 
$$
$$
Z_{10}(r_0,r_1,r_2,r_3)   =  -Z_{10}(r_0,-r_1-1,\pm r_2,\pm r_3)\,.
$$

\eject

\eject

\end{document}